\documentclass[twocolumn,prl,aps,epsf,showpacs,floatfix,superscriptaddress]{revtex4-1}

\usepackage{graphicx}
\usepackage{dcolumn}
\usepackage{bm}
\usepackage{hyperref}
\usepackage{amssymb}
\usepackage{xcolor}

\newcommand{\pienu}{$\pi^+ \rightarrow \mbox{e}^+ \nu$}

\newcommand{\pimue}{$\pi^+ \rightarrow \mu^+ \rightarrow \mbox{e}^+$}
\newcommand{\muenunu}{$\mu^+ \rightarrow \mbox{e}^+ \nu \overline{\nu}$}

\begin{document}

\preprint{APS/123-QED}

\title{Improved Search for Heavy Neutrinos in the Decay $\pi\rightarrow e\nu$}

\author{A.~Aguilar-Arevalo}
\affiliation{Instituto de Ciencias Nucleares, Universidad Nacional Aut\'onoma de M\'exico, CDMX 04510, M\'exico}

\author{ M.~Aoki}
\affiliation{Physics Department, Osaka University, Toyonaka, Osaka, 560-0043, Japan}

\author{M.~Blecher}
\affiliation{Virginia Tech., Blacksburg, VA, 24061, USA}

\author{D.I.~Britton}
\affiliation{SUPA - School of Physics and Astronomy, University of Glasgow, Glasgow, United Kingdom}

\author{D.~vom~Bruch}
\thanks{Present address: LPNHE, Sorbonne Universit\'e, Universit\'e Paris Diderot, CNRS/IN2P3, Paris, France}
\affiliation{Department of Physics and Astronomy, University of British Columbia, Vancouver, B.C., V6T 1Z1, Canada}

\author{D.A.~Bryman}
\affiliation{Department of Physics and Astronomy, University of British Columbia, Vancouver, B.C., V6T 1Z1, Canada}
\affiliation{TRIUMF, 4004 Wesbrook Mall, Vancouver, B.C., V6T 2A3, Canada}

\author{S.~Chen}
\affiliation{Department of Engineering Physics, Tsinghua University, Beijing, 100084, China}

\author{J.~Comfort}
\affiliation{Physics Department, Arizona State University, Tempe, AZ 85287, USA}

\author{S.~Cuen-Rochin}
\affiliation{Department of Physics and Astronomy, University of British Columbia, Vancouver, B.C., V6T 1Z1, Canada}

\author{L.~Doria}
\thanks{Corresponding author \\(luca@triumf.ca,~doria@uni-mainz.de).}
\affiliation{TRIUMF, 4004 Wesbrook Mall, Vancouver, B.C., V6T 2A3, Canada}
\affiliation{Institut f\"ur Kernphysik, Johannes Gutenberg-Universit\"at Mainz, Johann-Joachim-Becher-Weg 45, D 55128 Mainz, Germany}

\author{P.~Gumplinger}
\affiliation{TRIUMF, 4004 Wesbrook Mall, Vancouver, B.C., V6T 2A3, Canada}

\author{A.~Hussein}
\affiliation{TRIUMF, 4004 Wesbrook Mall, Vancouver, B.C., V6T 2A3, Canada}
\affiliation{University of Northern British Columbia, Prince George, B.C., V2N 4Z9, Canada}

\author{Y.~Igarashi}
\affiliation{KEK, 1-1 Oho, Tsukuba-shi, Ibaraki, Japan}

\author{S.~Ito}
\thanks{Present address: Faculty of Science, Okayama University, Okayama, 700-8530, Japan}
\affiliation{Physics Department, Osaka University, Toyonaka, Osaka, 560-0043, Japan}

\author{S.~Kettell}
\affiliation{Brookhaven National Laboratory, Upton, NY, 11973-5000, USA}

\author{L.~Kurchaninov}
\affiliation{TRIUMF, 4004 Wesbrook Mall, Vancouver, B.C., V6T 2A3, Canada}

\author{L.S.~Littenberg}
\affiliation{Brookhaven National Laboratory, Upton, NY, 11973-5000, USA}

\author{C.~Malbrunot}
\thanks{Present address: Experimental Physics Department, CERN, Gen\`eve 23, CH-1211, Switzerland}
\affiliation{Department of Physics and Astronomy, University of British Columbia, Vancouver, B.C., V6T 1Z1, Canada}

\author{R.E.~Mischke}
\affiliation{TRIUMF, 4004 Wesbrook Mall, Vancouver, B.C., V6T 2A3, Canada}

\author{T.~Numao}
\affiliation{TRIUMF, 4004 Wesbrook Mall, Vancouver, B.C., V6T 2A3, Canada}

\author{D.~Protopopescu}
\affiliation{SUPA - School of Physics and Astronomy, University of Glasgow, Glasgow, United Kingdom}

\author{A.~Sher}
\affiliation{TRIUMF, 4004 Wesbrook Mall, Vancouver, B.C., V6T 2A3, Canada}

\author{T.~Sullivan}
\thanks{Present address: Department of Physics, Queen’s University, Kingston K7L 3N6, Canada}
\affiliation{Department of Physics and Astronomy, University of British Columbia, Vancouver, B.C., V6T 1Z1, Canada}

\author{D.~Vavilov}
\affiliation{TRIUMF, 4004 Wesbrook Mall, Vancouver, B.C., V6T 2A3, Canada}


\collaboration{PIENU Collaboration}
  
\date{\today}
             
\begin{abstract}
  A search for massive neutrinos has been made in the decay \pienu.  
  No evidence was found for extra peaks in the positron energy spectrum
  indicative of pion decays involving massive neutrinos ($\pi\rightarrow e^+ \nu_h$).
  Upper limits (90 \% C.L.) on the neutrino mixing matrix element $|U_{ei}|^2$
  in the neutrino mass region 60--135 MeV/$c^2$ were set, which are 
  an order of magnitude improvement over previous results.
\end{abstract}

\maketitle

\section{Introduction}
In the original Standard Model (SM) \cite{sm}, neutrinos are included as massless particles.
There is now firm experimental evidence that neutrinos oscillate between different flavors,
indicating that at least two are massive particles~\cite{deg}.
Many extensions of the SM incorporating massive neutrinos hypothesize the existence
of additional neutrino states.
Right-handed gauge-singlets (sterile neutrinos), are an essential ingredient
in seesaw models~\cite{ss} aiming to explain the smallness of neutrino masses.
In the Neutrino Minimal Standard Model~\cite{numsm} ($\nu$MSM) three sterile neutrinos
and three corresponding mass eigenstates are added, leading to new mixings between
six definite mass states and the active and sterile states.
For example, depending on the choices of parameters and mass hierarchy in the $\nu$MSM,
the two heaviest sterile neutrino states may occur in the range probed by meson decays \cite{aei},
while the lightest state can play a role as a dark matter particle in the keV/c$^2$ mass range.
Massive neutrinos in the MeV/c$^2$ range are also required in dark matter models addressing 
small scale structure problems \cite{mckeen1} or involving  new thermalization scenarios \cite{mckeen2}.
More generally, for $k$ sterile neutrinos, the weak eigenstates $\nu_{\chi_k}$ are related to the
mass eigenstates $\nu_i$ by a unitary transformation matrix $U_{\ell i}$, where
$\nu_{\ell} =\Sigma_{i=1}^{3+k} U_{\ell i} \nu_i$,
with $\ell = e, \mu, \tau, \chi_1, \chi_2...\chi_k$.

Depending on the mass scale of the new heavy mass eigenstates, sterile neutrinos
can have different phenomenological signatures.
If heavy neutrino states are Majorana fermions, neutrinoless double
beta decay experiments may provide stringent constraints~\cite{gouvea}.
Other constraints on $U_{\ell i}$ come from lepton universality tests, the decay width of invisible decays of Z bosons,
$\mu$ and $\tau$ lepton-flavor-violating decays, and magnetic and electric dipole moments of charged leptons~\cite{gouvea}.
In particular, heavy neutrinos
with MeV/c$^2$ to GeV/c$^2$ masses can have measurable effects in meson decays that
can be explored by precisely measuring their decay branching ratios or
by searching for extra peaks in the energy spectrum of their leptonic two-body decays 
(e.g. $\pi,K,B\rightarrow l\nu$)~\cite{shrock}.

The decay \pienu~ (positron energy $E_{e^+} = 69.8$ MeV) is helicity-suppressed in the SM 
and its measured branching ratio is $R_{\textrm{exp}}=(1.2327 \pm 0.0023) \times 10^{-4}$ \cite{pdg,triumf,psi,pienuprl}.
The presence of heavy neutrinos relaxes the helicity suppression; comparing the experimental
value with the theoretical SM calculation
$R_{SM}=(1.2352 \pm 0.0002)\times 10^{-4}$~\cite{marciano,finke,vincenzo}, limits on
$|U_{ei}|^2$ have been obtained for masses below 60~MeV/$c^2$ \cite{pienuprl}.
A previous search for additional peaks in \pienu~ decays~\cite{leener,oldneutrino} established upper limits at the
level of $|U_{ei}|^2 < 10^{-7}$ in the neutrino mass region of
50--130 MeV/$c^2$ and was limited by the presence of the \muenunu~ decay
background ($E_{e^+} = 0.5 - 52.8$ MeV) originating from decays-in-flight of pions.
An improved limit was published in Ref.~\cite{mnu} based on a partial data set from the PIENU experiment.

In this work, we present a search for additional peaks in the low-energy region of the
background-suppressed \pienu~ spectrum using the full data set collected by the PIENU
experiment, representing a sample larger than \cite{mnu} by an order of magnitude.
\begin{figure}
  \includegraphics[scale=0.7]{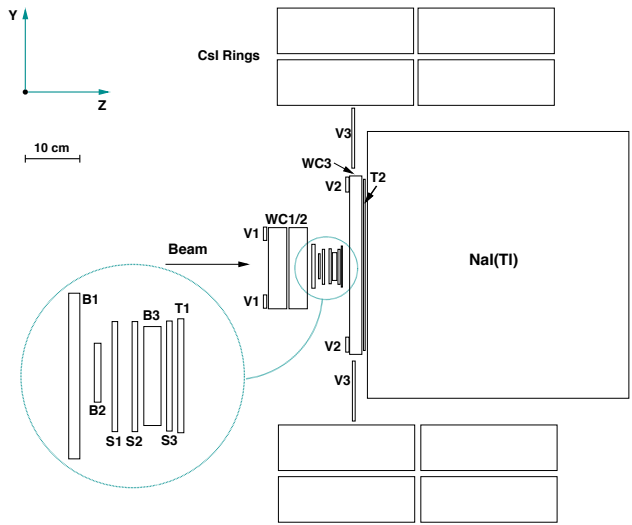}%
  \caption{\label{fig:DET}Schematic illustration of the PIENU detectors with the pion stopping region
    shown in the inset.}
\end{figure}
\section{Experimental Technique}
The $\pi^+$ beam was provided by the TRIUMF M13 beamline, modified
to deliver $75\pm 1$ MeV/$c$ pions with $<2$\% positron contamination~\cite{m13}.
A detailed description of the PIENU detector (Fig. \ref{fig:DET}) can be found in~\cite{pienudetector}.
Briefly, beam tracking was realized by two sets of multiwire proportional chambers (WC1 and 2), each with three
planes of wires oriented at 60$^{\circ}$ to each other. After WC1/2, the beam was degraded by
two plastic scintillators (B1 and B2). These provided the pion arrival trigger signal, energy loss measurement
for particle identification, and detection of extra beam particles.
Pions were stopped in an 8~mm thick plastic scintillator (B3) where they decayed at rest via
\pienu~ or \pimue~ ($\pi^+ \rightarrow \mu^+\nu$ followed by $\mu^+ \rightarrow \mbox{e}^+ \nu \overline{\nu}$).
Two sets of silicon microstrip detectors (S1, S2), each with two orthogonal planes, were installed upstream
of B3 for tracking pions and (together with the pion track information from WC1/2) to detect decays-in-flight.
After B3, a third silicon microstrip detector (S3)
and a wire chamber (WC3) provided tracking for decay positrons.

Two plastic scintillators (T1 and T2) provided the decay trigger signal.
WC3 and T2 were mounted in front of a NaI($T\ell$) calorimeter crystal (a 48~cm diam. $\times$~48~cm long cylinder),
which provided the main positron energy measurement.
The NaI($T\ell$) calorimeter was surrounded by 97 pure CsI crystals arranged in a two-layer concentric 
structure for electromagnetic shower leakage detection.
The trigger logic was based on a pion signal provided by the coincidence B1$\cdot$B2$\cdot$B3
(with a high B1 threshold to select pions) and a decay-positron signal provided by a T1$\cdot$T2 coincidence.
A coincidence of pion-arrival and decay-positron signals within a time window from -300~ns to 540~ns
defined the logic for an unbiased collection of events, prescaled by a factor 16.
Another trigger was based on a decay-positron signal in the 2~ns to 40~ns time window without prescaling,
containing most of the \pienu~events.
Continuous calibration of the detectors was provided by dedicated beam-positron and cosmic-ray triggers.
The typical pion stopping rate was $5 \cdot 10^{4}$~s$^{-1}$, while the trigger rate was 600~s$^{-1}$.
Plastic scintillators, NaI(Tl) and CsI calorimeters were read out by waveform digitizers at
 500, 30, and 60~MHz respectively. Silicon detectors were digitized at 60~MHz, while the wire chambers and
trigger signals were read by multi-hit TDCs with 0.625~ns resolution.

\section{Data Analysis}
\subsection{Event Selection}
The analysis included four data sets with approximately $10^7$ \pienu~ events collected over a period of four years,
including the data used in the previous search~\cite{mnu} based on $10^6$ \pienu~ events.
Automatic calibration procedures and run-by-run gain stability corrections
of the detectors ensured similar decay spectra for all data sets. 
The event selection adopted the same strategy as for the extraction of the \pienu~ branching ratio~\cite{pienuprl}.
Pions were identified based on their energy loss in B1 and B2 and a cut was applied to WC1/2 to exclude beam halo particles.
Cuts based on waveform information from B1, B2 and T1 were used to reject events with additional beam particles.
\begin{figure*}
  \includegraphics[scale=0.85]{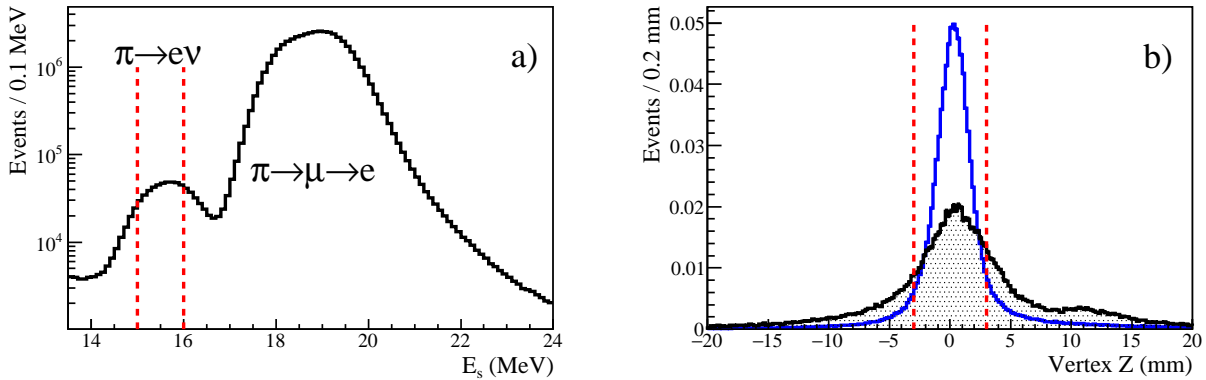}%
  \caption{\label{fig:vars} {\bf a)} Energy sum $E_s$ measured in B1, B2, S1, S2, and B3.
    {\bf b)} Z vertex for events with positron energy $E_{e^+}<52$~MeV
    (shaded histogram) and $E_{e^+}>52$~MeV (blue full line). The two distributions are normalized to contain the same number of events
    and cuts applied are indicated by the red vertical dashed lines.}
\end{figure*}
A fiducial radial cut of 80~mm in WC3 was used,
resulting in 20\% solid angle acceptance for positron tracks.
A requirement of $<2$~MeV measured by the CsI array was applied
to select events which were mainly contained in the higher resolution NaI($T\ell$) detector.

Suppression of \pimue~ events was based on timing, energy, and tracking information.
Events in the 4--35~ns timing window were selected. 
The strongest suppression factor was given by the sum $E_s$ of the energy deposits in B1, B2, S1, S2,
and B3 (with 100~ns integration window).
For \pimue~ decay, the energy deposit in B3 was generally larger than for \pienu~ decay
by 4.1~MeV, due to the presence of the muon.
However, due to the 100~ns integration window which might miss the positron and light output saturation
effects in the plastic scintillator, the energy distribution observed in B3 for \pimue events overlapped that for \pienu~events.
Therefore, a cut in $E_{s}$ with a width of 1~MeV was applied, as indicated in Fig.~\ref{fig:vars}a,
to minimize the \pimue background.

The beam tracking detectors WC1/2 and S1/2 allowed the measurement of the vertex of pion decays-in-flight
before B3 and, when combined with positron tracking information from S3 and WC3 downstream of B3, allowed
an estimate of the decay vertex in the beam direction (Z); this latter distribution is 
broader in the case of \pimue~ events due to the distance traveled by the muon in B3.
Cuts on the pion decay-in-flight track angle upstream of B3~\cite{mnu} and Z
vertex distributions (Fig.~\ref{fig:vars}b) helped in rejecting the \pimue~ backgrounds.

The suppression cuts were optimized to minimize the figure of merit $\sqrt{N_{L}}/N_{H}$ where $N_{L}$
and $N_{H}$ are the numbers of events below and above 52~MeV in the positron energy spectrum.
The cuts suppressed the backgrounds by a factor of $10^7$ with the final positron energy spectrum, yielding
$\sqrt{N_{L}}/N_{H} = 2.8\cdot 10^{-4}$. 
The background-suppressed spectrum is shown in Fig.~\ref{fig:FIT}, where the majority of events are of the \pienu~ type concentrated
at the peak at 69.8~MeV with a low-energy tail extending below the background events (mainly \pimue~ where the pion or the muon
decayed in flight near or in B3). The shoulder at about 58~MeV is due to photonuclear reactions in the NaI($T\ell$)
followed by neutron emission and escape from the crystal~\cite{nai}.

\subsection{Positron Spectrum Fit}
\begin{figure*}
  \includegraphics[scale=0.8]{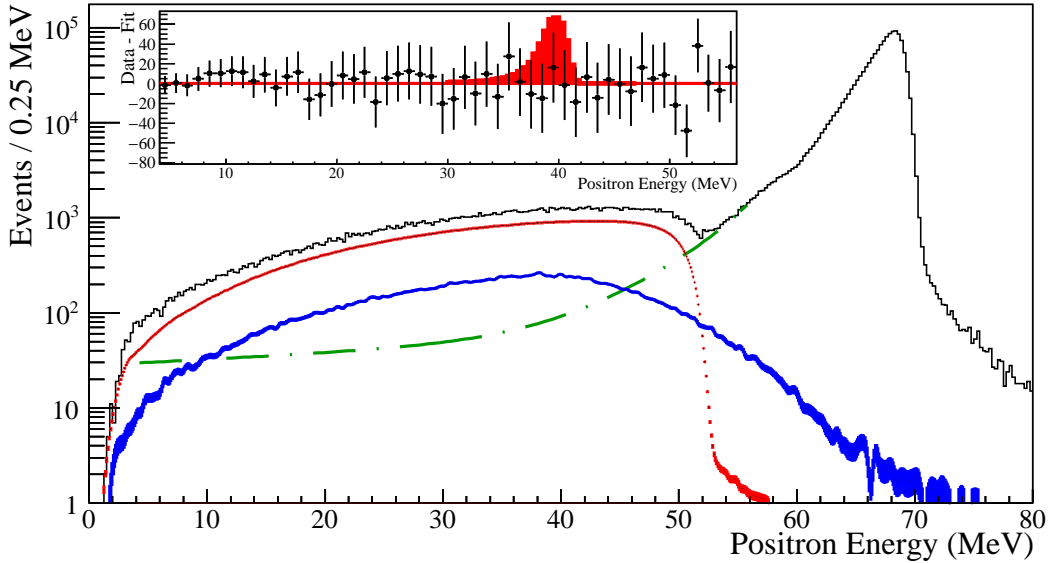}
  \caption{\label{fig:FIT}Background-suppressed positron energy spectrum (black histogram).
    Fitted components include muon decays in flight (thick blue line, from MC), \pienu~ (green, dot-dashed line, fit to MC),
    and \pimue~ (red dashed line, from late-time data events).
    The insert shows the (rebinned) residuals (Data--Fit) with statistical error bars and
      the signal shape in the case of $E_{e^+}=40$~MeV and $|U_{ei}|^2=10^{-8}$.
  } 
\end{figure*}

The background-suppressed positron spectrum was used to search for additional peaks due to massive neutrinos.
The spectrum was fitted from $E_{e^+}=4$~MeV to $E_{e^+}=56$~MeV with background components and a
signal component. 
The lower limit of the fit is set by the lack of statistics and the sharp drop of the cut efficiencies,
while the upper limit is set for avoiding the peaks due to photonuclear interactions.
The fit was repeated shifting the signal component in 0.25~MeV steps within the fitting range.
The signal shape was simulated for every energy step with a Monte Carlo (MC) simulation validated with an experimental
study of the calorimeter response to positrons.
The shapes of the background components were obtained from late-time positrons ($t>200$~ns) representing the \pimue~ decay chain,
a component derived from MC describing \pimue~ events where the muon decayed in flight in B3, mimicking the \pienu~ timing, and
a \pienu~ low-energy tail component due to electromagnetic shower losses, which was a triple-exponential fit to MC spectrum.
The fitted components are shown in Fig.~\ref{fig:FIT}.
The background-only fit described the data well, yielding $\chi^2/\textrm{dof} = 197.2/203 = 0.97$ and the
addition of purported signals did not change the result.
Since no significant excesses beyond statistical fluctuations were found, upper 
limits $|U_{ei}|^2_{UL}$  on the couplings could be calculated.

\subsection{Acceptance Correction}
\begin{figure}
  \includegraphics[scale=0.45]{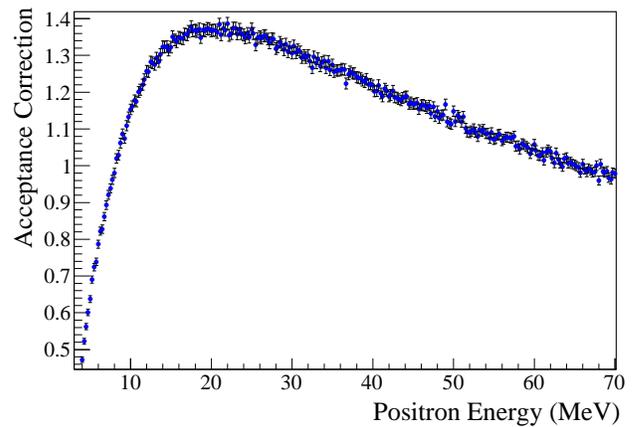}
  \caption{Acceptance correction $Acc(E_{e^+})$ as a function of the positron energy as determined with the MC simulation.
  The error bars are statistical only.}
  \label{fig:ACC}
\end{figure}

Most systematic and acceptance effects canceled to first order, since the upper limit is proportional to the
ratio of the fitted number of signal events and the number of \pienu~ events $N(\pi\rightarrow e\nu_e)$,
which were estimated by fitting a MC generated spectrum to the data for $E_{e^+}>52$~MeV.
However, energy-dependent effects induced by the suppression cuts did not completely cancel,
and a correction was needed.
The positron energy-dependent acceptance correction $Acc(E_{e^+})$ was calculated via a MC simulation.
Uniformly distributed positron tracks were simulated at a given energy $E_{e^+}$ between 0 and 70~MeV
in 0.25~MeV steps with the suppression cuts: energy sum $E_s$, Z vertex, and CsI veto.
The ratio between the number of events at a given $E_{e^+}$ and the number of events at $E_{e^+}=70$ MeV
was taken as the relative acceptance correction.
The correction as a function of the positron energy is shown in Fig.~\ref{fig:ACC}. The statistical uncertainty
due to the MC procedure is about 1\%. The increase of the acceptance correction towards low energies is due to the
CsI veto cut, which removes more high-energy events having larger shower leakage from the NaI(Tl) calorimeter.
In order to estimate the systematic uncertainty on the acceptance correction, a study was performed
using \pimue~ events. In contrast to \pienu~ events, the background contribution to this spectrum was negligible;
it has higher statistics and covers a broad energy range.
Comparing \pimue~ events to the MC, the effect of the suppression cuts was studied and good agreement
with the data was found.
The maximum difference between data and MC for $E_{e^+}>10$ MeV was 3\%, which was conservatively
assigned to be the systematic uncertainty on the acceptance correction.

\begin{figure*}
  \includegraphics[scale=0.9]{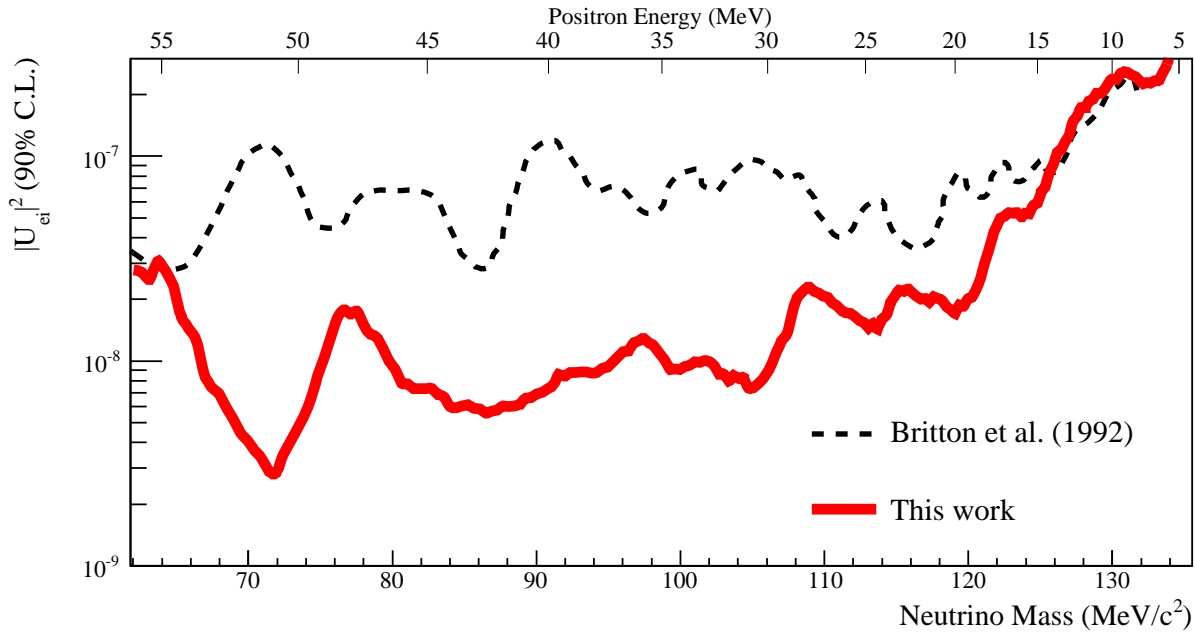} 
  \caption{\label{fig:UL} 90\% C.L. upper limits on the square of the mixing matrix elements $|U_{ei}|^2$
    of heavy neutrinos coupled to electrons (thick red line).
    The black dashed line shows the results from~\cite{oldneutrino}.
  }
\end{figure*}

\section{Results}
Since no significant peaks were found in the data, 90\% C.L. upper limits $N(\pi\rightarrow e\nu_i)_{UL}$ were
calculated with a Bayesian
procedure, assuming a flat prior and enforcing a positive peak amplitude and a Gaussian probability distribution.

An upper limit $|U_{ei}|^2_{UL}$ on the squared matrix element describing the mixing of the massive states with
the other active neutrino states was obtained using
\begin{equation}
  \frac{1}{Acc(E_{e^+})} \frac{N(\pi\rightarrow e\nu_i)_{UL}}{N(\pi\rightarrow e\nu)}=|U_{ei}|^2_{UL} \rho_e(E_{e^+}) \quad,
  \label{limit}
\end{equation}
where $\rho_e(E_{e^+})$ is a phase space and helicity-suppression factor~\cite{shrock}
\begin{eqnarray}
  \rho_e(E_{e^+}) = \frac{\sqrt{1+\delta_e^2+\delta_i^2-2(\delta_e+\delta_i+\delta_e\delta_i)}}
      {\delta_e(1-\delta_e)^2}\nonumber \\
      \times (\delta_e+\delta_i-(\delta_e-\delta_i)^2) \quad,
\end{eqnarray}
where
\[
\delta_e=m_e/m_{\pi} \quad , \quad \delta_i=m_{\nu_i}/m_{\pi} \quad \textrm{, and} 
\]
\[
 m_{\nu_i}=\sqrt{m_{\pi}^2-2m_{\pi}E_{e^+}+m_e^2} \quad.
\]

The results for the 90\% C.L. upper limits for $|U_{ei}|^2$
are shown in Fig.~\ref{fig:UL} (thick red line), together with
the previous result~\cite{oldneutrino} (black dashed line).
These results supersede those reported in \cite{mnu}.

\section{Summary}
A search has been performed for the mixing of heavy neutrinos coupled to electrons in the decay $\pi^+\rightarrow e^+\nu_h$.
No extra peaks due to heavy neutrinos were found in the positron energy spectrum, resulting in 
upper limits set on the square of the mixing matrix elements $|U_{ei}|^2$ from $10^{-8}$ to $10^{-7}$
for neutrino masses in the range 60--135 MeV/$c^2$. These results are independent of assumptions about the
nature of the heavy neutrino and are comparable to limits from neutrinoless double beta decay found in~\cite{gouvea},
which assume that massive neutrinos are Majorana in nature.\\

\begin{acknowledgments}
This work was supported by the Natural Sciences and Engineering Research Council of Canada and TRIUMF through a contribution
from the National Research Council of Canada, and by Research Fund for Doctoral Program of Higher Education of China, and
partially supported by KAKENHI (18540274, 21340059) in Japan.
One of the authors (M.B.) was supported by US National Science Foundation Grant Phy-0553611.
We are indebted to Brookhaven National Laboratory for the loan of the crystals.
We would like to thank the TRIUMF detector, electronics and DAQ groups for the extensive support.
We would like to thank A. de Gouv\^ea and A. Kobach for providing useful information.
\end{acknowledgments}

\end{document}